\documentclass{elsart}

\usepackage{natbib}

\usepackage{graphicx}
 
\renewcommand\deg{\ifmmode^\circ\else$^\circ$\fi}

\begin{document}


\runauthor{Kraus, Witzel, \& Krichbaum}


\begin{frontmatter} 
\title{Intraday Radio Variability in Active Galactic Nuclei}
\author[MPIfR]{A. Kraus}
\author[MPIfR]{A. Witzel}
\author[MPIfR]{T.P. Krichbaum}

\address[MPIfR]{Max-Planck-Institut f\"ur Radioastronomie, Auf dem
H\"ugel 69, 53121 Bonn, Germany}


\begin{abstract} 
  
Rapid flux density variations on timescales of the order of a day or
less (Intraday Variability, IDV) in the radio regime are a common
phenomenon within the blazar class.  Observations with the 100m
telescope of the MPIfR showed that the variations occur not only in
total intensity, but also in the polarized intensity and in
polarization angle.  Here we present an overview of our
IDV-observations and discuss briefly some models which
may explain the effect.

\end{abstract} 



\end{frontmatter} 

\section{Introduction}
\label{intro} 

Flux density variations of extragalactic radio sources on timescales
of the order of several weeks to years are well-known since the
mid-sixties (e.g.\ \citet{Kell1} and references therein).  They are
used to study the physics of AGN, and have led -- together with early VLBI
results -- to the development of the relativistic jet-model.  In 1985,
observations with the 100m telescope of the MPIfR in Effelsberg
detected significantly faster intensity variations (on timescales of a
few days down to several hours), the so-called {\bf I}ntra{\bf D}ay
{\bf V}ariability (IDV) \citep{Witz1,Hees1}.  These rapid variations
were studied in some detail in the following years, and it turned out
that they are quite common in compact extragalactic radio sources.
Recently, IDV was also discovered in sources in the southern
hemisphere \citep{Ked1}.

\section{Observations and Results} 
\label{obs} 

Our observations of IDV in AGN have been carried out at the 100m
telescope of the MPIfR in Effelsberg and at the VLA, studying
variations of the total flux density, and --- more recently --- also
of the (linear) polarization.  For the total intensity, elevation- and
time-dependent effects have been corrected using steep-spectrum
sources, which do not show any IDV.  For the polarization
observations, we correct the instrumental polarization and the
``cross-talk'' between the Stokes-channels, applying the Matrix-Method
proposed by \citet{Tur1}.  With these procedures, we are able to reach
relative measurement errors of 0.3--1.2\,\% (depending on the
wavelength and the weather conditions) for the total flux density,
3--5\,\% for the polarized flux density, and 2--5$\deg$\/ for the
polarization angle (for the rare highly polarized sources the
measurement errors of the latter two quantities can be somewhat
smaller).

\begin{figure}[hbt] 
\centering
\vspace*{0.2cm}
\includegraphics[scale=0.52]{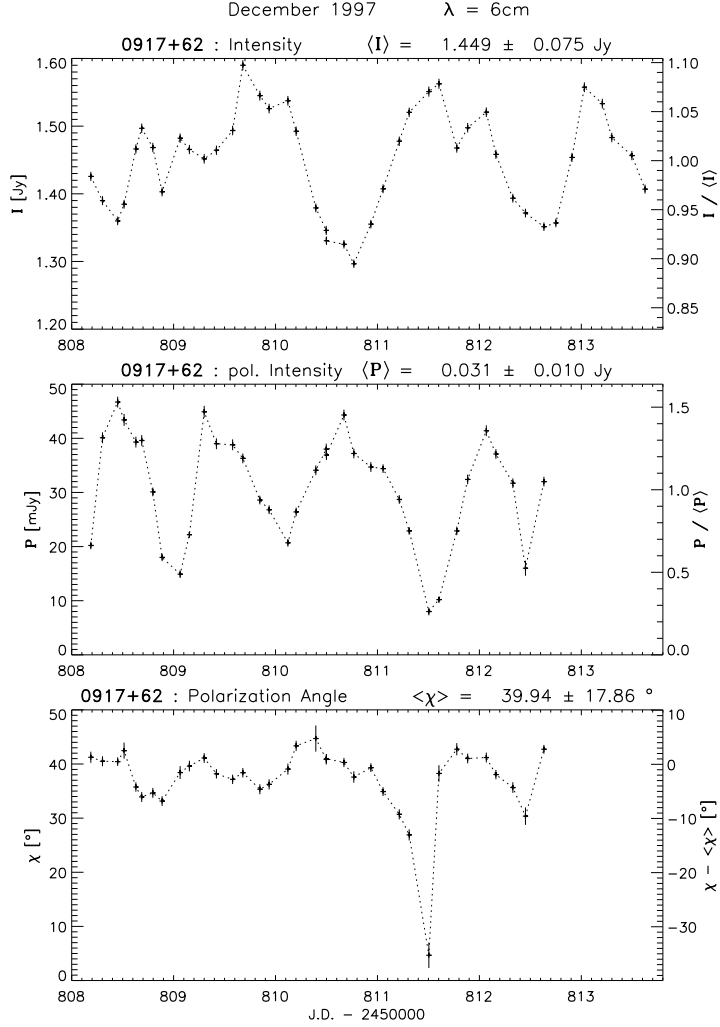}
\caption{IDV in the Quasar 0917+624, observed in December 1997 at
$\lambda=6$\,cm. We plotted the total flux density $I$ (upper panel), the
polarized flux density $P$ (in the middle) and the polarization angle
$\chi$ (lower panel) against Julian date. The variations in total and
polarized flux density seem to be anti-correlated (see text and Fig.\
\ref{fig2}). At the end of the observation we could not achieve
polarization data due to technical problems.}
\label{fig1}
\end{figure}

Since 1985, we have observed 73 AGN (some repeatedly) in search for
IDV; this includes the complete subsample of flat-spectrum sources of
the 1-Jy-catalog north of $\delta = 50$\deg. It turned out that the
rapid variability is a common phenomenon in compact flat-spectrum
radio sources: one third of the observed sources show variations with
timescales of $\leq 2$ days, one third show variability on longer
timescales, and only one third of those sources never showed short
timescale variations.
\begin{figure}[hbt] 
\centering
\includegraphics[scale=0.36,angle=90]{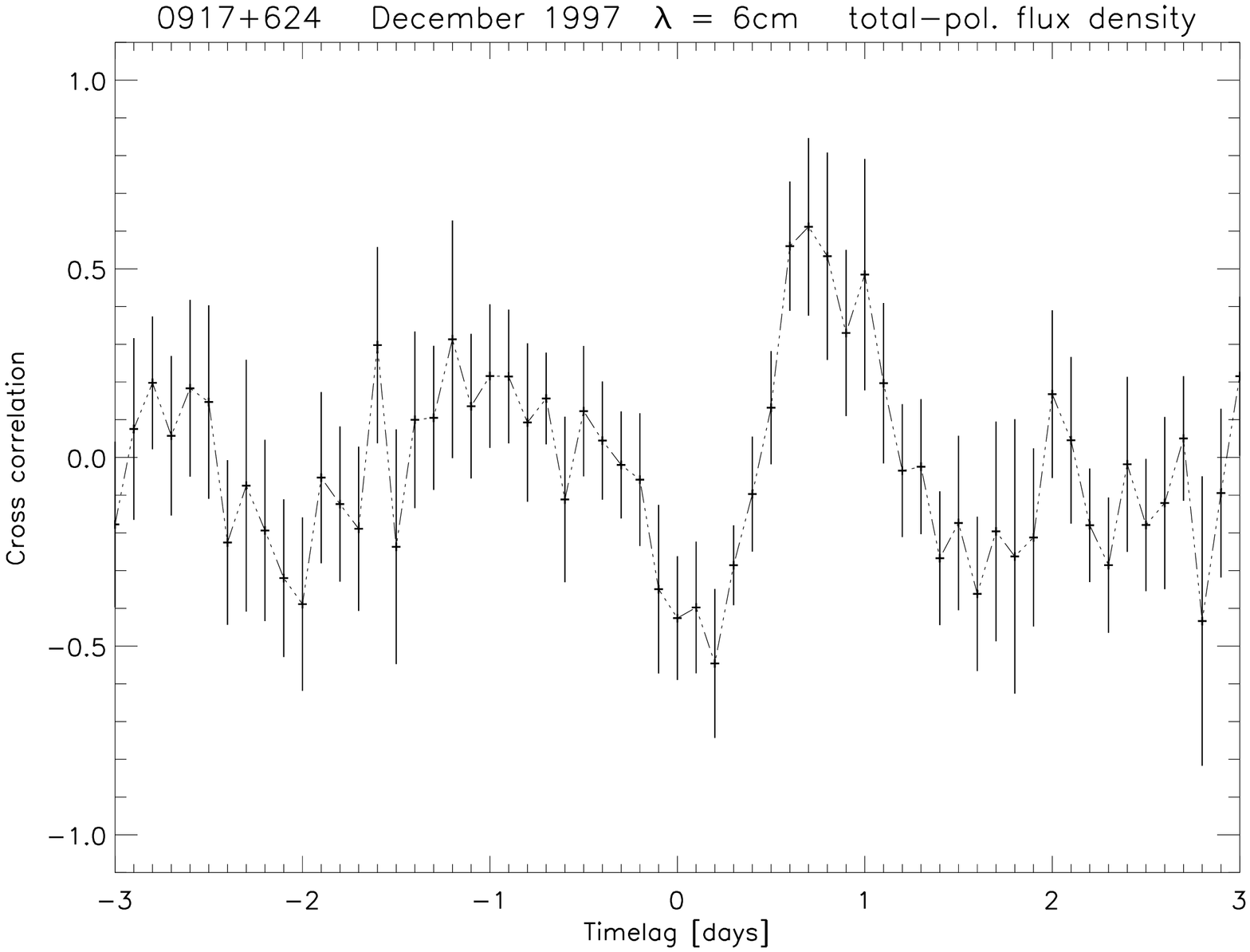}
\caption{Cross-Correlation analysis for the total and the polarized
intensity of the light curve seen in Fig.\ \ref{fig1}. We plotted the
Cross-Correlation Function (CCF) between $I$ and $P$ against the
timelag $\Delta\tau$. This confirms the anti-correlation between both
data series (minimum at $\Delta\tau \simeq 0$). The CCF reveals
further that the total intensity precedes the polarized flux density
by about $\Delta\tau \simeq 0.7$\,days (see text).}
\label{fig2}
\end{figure}
Furthermore, the observations revealed that IDV is not only present in
total intensity $I$, but is usually accompanied by variability of the
linear polarization (intensity $P$ and position angle $\chi$).  While
total intensity variations range from a few percent up to 35\,\%
(e.g.\ in the case of the QSO 0804+499, \citet{Qui2}), variability in
the polarized intensity is usually larger and can reach a factor of
two e.g.\ in the QSO 0917+624 \citep{Kraus1}.  So far, we have found
no significant correlation between the strength of the variations or
the timescales with either the redshift of the source, the galactic
latitude, or the spectral index.

In Fig.\ \ref{fig1}, we show, as an example of the rapid variations,
the variability observed in the quasar 0917+624 in total and polarized
flux density and polarization angle (from top to bottom).  An
anti-correlation between the total and the polarized flux density is
clearly present in the light curves. This can be supported further by
the computation of the Cross-Correlation Function (CCF, e.g.\
\citet{edelson}) between $I$ and $P$ which is plotted in Fig.\
\ref{fig2}. The minimum close to the timelag $\Delta \tau =0$
confirmes this anti-correlation. Due to the quasi-periodicity of both
light curves, the polarized flux density variability seems to be
phase-shifted with respect to the total intensity variations. This
corresponds to the maximum of the CCF at $\Delta\tau\simeq 0.7$\,days.
The anti-correlation between $I$ and $P$ is seen frequently in
0917+624, while other sources (e.g.\ the BL\,Lac object 0716+714)
rather show a direct correlation between both values (e.g.\
\citet{Wagn1}, Kraus et al., in preparation).  Similar polarization
angle variability was also observed in other sources. In 0917+624 the
angular variations usually are larger than shown in Fig.\ \ref{fig1}
(e.g.\ \citet{Kraus1}).  Once, even a 180\deg-swing was observed
\citep{Qui3}.

\begin{figure}[hbt] 
\centering
\includegraphics[scale=0.36,angle=90]{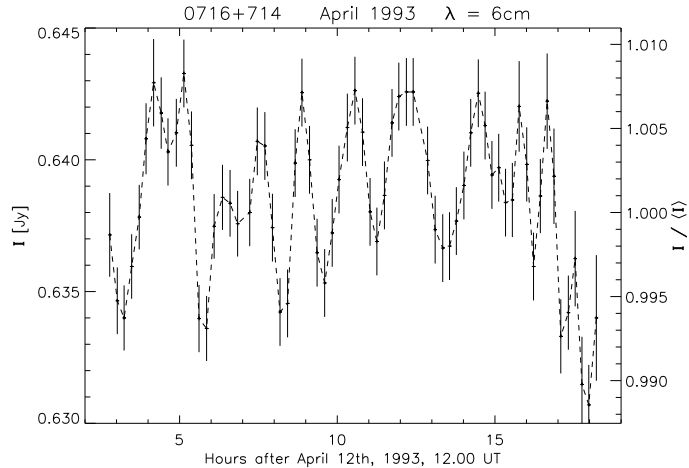}
\caption{Rapid variability in the BL\,Lac object 0716+714.  We plotted
the flux density against the time in hours. Obviously, there are
significant variations on timescales of about 1--2 hours, although on
a low percentage level. This is the shortest timescale seen in our
experiments.}
\label{fig4}
\end{figure}

In the BL\,Lac object 0716+714, we observed a direct correlation
between the radio and the optical flux density variations
\citep{Qui1}, and discovered in April 1993 even faster variations (on
timescales of two hours) than in any other source before (Fig.\
\ref{fig4}).  It is unclear whether such rapid variability occured
only in this source, or has not been found before because of
undersampling in time.

\section{Discussion and Conclusions}
\label{theory}

Thirteen years after its discovery, IDV is still not fully understood.
In the case of an intrinsic origin, the timescale of the variations
corresponds directly to the size of the emitting region, via the light
travel time relation ($R \simeq c \cdot t_{obs}$).  From this and the
Rayleigh-Jeans law the brightness temperature of the variable
component can be derived \citep{Wagn1}, resulting in values for $T_B$
in the range of $10^{17}$--$10^{21}$\,K, far in excess of the inverse
Compton limit of $10^{12}$\,K \citep{Kell2}.  This fact seriously
challenges existing models proposed to explain AGN variability.  On
the other hand, assuming an intrinsic origin of IDV, the investigation
of IDV offers a method to study the physical properties of AGN on very
small scales (in the order of light days or even smaller).

A correlation between the variations in the radio and the optical
bands (seen e.g.\ in 0716+714 by \citet{Qui1}) argues in favour of an
intrinsic origin of IDV.  In addition, the lack of a clear dependence
of the strength or the timescale of IDV on the frequency or the
galactic latitude speaks against interstellar scattering (ISS,
\citet{Rick1}) as the exclusive cause of IDV.  (Nevertheless, owing to
the small source sizes $R \simeq c \cdot t_{obs}$ involved, ISS should
be present as additional effect in the radio band.)  Gravitational
microlensing as an alternative extrinsic explanation is implausible
because of the high duty cycle and short timescales of the variations
and the fact that source sizes of the order of tens of $\mu$as are
needed \citep{Wagn1}.

It is clear, that the variations of the linear polarization,
especially the polarization angle variations, require special
models. The easiest assumption might be a model with two or more
independent components, taking into account the vector addition for
the polarization vector. In fact, for extrinsic explanations like ISS
or microlensing, this scheme is inevitable (e.g.\ \citet{Wagn1}).  In
the light of the shock-in-jet models, \citet{Koen1} explain changes of
the polarization angle by the successive illumination of
cross-sections with different magnetic field orientations in the
jet. In the case of a small viewing angle of the jet, even
180\deg-swings (as observed in 0917+624 by \citet{Qui3}) are
possible. In an alternative model (Qian et al., in preparation), a
thin sheet of relativistic electrons moves along magnetic field lines
with a very high Lorentz factor ($\Gamma \simeq 20$--25). The observed
variability is then explained by minor changes of the viewing angle
(by only a few degrees) which give rise to large variations of the
aberration angle and, therefore, of the observed synchrotron emission.

Adopting an intrinsic mechanism for IDV, \citet{Qian1,Qian2}
considered the propagation of a thin shock through the jet plasma in a
cylindrical geometry with periodic boundaries.  They found this model
capable of explaining the variations and the apparent high brightness
temperatures by $T_B^{\rm app} = \gamma^2_s \, \delta^3 \, T_B^{\rm
true}$.  Therefore, the high brightness temperatures can be reached
easier, although even in this case Doppler factors higher than usually
observed are needed.  Recently, \citet{Spada} discussed a model in
which the radiating electrons are accelerated by shocks in a conical
geometry.  If the injection times are shorter than the variability
timescale, brightness temperatures of up to $10^{17}$\,K can be
explained with moderate Lorentz factors ($\Gamma \simeq 10$).
Alternatively, collective emission processes proposed e.g.\ by
\citet{Benf1} can avoid the violation of the inverse Compton limit.
At present, however, it is unclear whether this process can produce
correlated broad-band (i.e., radio-optical) variations.

Thus, coordinated multifrequency observing campaigns covering a large
range of the electromagnetic spectrum are needed to distinguish
between the various models proposed.

{\it We thank A.P. Lobanov and E. Ros for critically reading the manuscript.}

\end{document}